\documentclass[12pt]{article}
\usepackage{amsmath,latexsym,amssymb,epsfig}%,psfrag}
\usepackage[varg]{txfonts}   % Web of Conferences font
\newcommand{\al}{\ensuremath{\alpha}}

\newcommand{\si}{\ensuremath{\sigma}}

\newcommand{\ep}{\ensuremath{\epsilon}}

\newcommand{\tr}{\mathop{\rm tr}}
\newcommand{\Tr}{\mathop{\rm Tr}}

\newcommand{\Group}[2]{{\hbox{{\itshape{#1}}($#2$)}}}
\newcommand{\U}[1]{\Group{U\kern0.05em}{#1}}
\newcommand{\SU}[1]{\Group{SU\kern0.1em}{#1}}
\newcommand{\SL}[1]{\Group{SL\kern0.05em}{#1}}
\newcommand{\Sp}[1]{\Group{Sp\kern0.05em}{#1}}
\newcommand{\SO}[1]{\Group{SO\kern0.1em}{#1}}

\newcommand{\scr}[1]{\ensuremath{\mathcal{#1}}}

\newcommand{\mybar}[1]%
	{{\kern 0.8pt\overline{\kern -0.8pt#1\kern -0.8pt}\kern 0.8pt}}
\newcommand{\sla}[1]%
	{{\raise.15ex\hbox{$/$}\kern-.57em #1}}% Feynman slash
\newcommand{\roughly}[1]%
	{{\mathrel{\raise.3ex\hbox{$#1$\kern-.75em\lower1ex\hbox{$\sim$}}}}}

\newcommand{\nop}[1]{:\kern-.3em#1\kern-.3em:}

\newcommand{\del}{\partial}

\newcommand{\beq}{\begin{equation}}
\newcommand{\eeq}{\end{equation}}
\newcommand{\bea}{\begin{eqnarray}}
\newcommand{\eea}{\end{eqnarray}}
\newcommand{\half}{\frac{1}{2}}
\newcommand{\p}{\partial}

\newcommand{\bep}{\bar{\epsilon}}
\def\a{\dot{\alpha}}
\def\ee{\varepsilon}
\newcommand{\dslash}{\slash{\!\!\!\del}}

\newcommand{\beqn}{\begin{eqalignno*}}
\newcommand{\eeqn}{\end{eqalignno*}}

\usepackage{graphicx} 

\begin{document}
\begin{titlepage}

\vskip 2cm
\begin{center} 
\vspace{0.5cm} \Large {\sc (Pseudo)goldstinos, SUSY fluids, Dirac gravitino and  gauginos}
\vspace*{5mm} 
\normalsize

\vspace*{5mm} 

{\bf
Karim~Benakli
\footnote{kbenakli@lpthe.jussieu.fr}
}
\vspace*{5mm} 

\emph{Sorbonne Universit\'es, UPMC Univ Paris 06, UMR 7589, LPTHE, F-75005, Paris, France \\
and \\
 CNRS, UMR 7589, LPTHE, F-75005, Paris, France}
\end{center}
\vskip0.6in 

\abstract{%
 We review the emergence and fate of goldstinos in different frameworks. First, we consider a super-Higgs mechanism
when supersymmetry breaking is induced by neither an F-term nor a D-term but related to a more general stress energy-momentum tensor. 
This allows us to build a novel Lagrangian that describes the propagation of a spin-{$\frac{3}{2}$} state in a fluid. 
Then we briefly review the ubiquitous pseudo-goldstinos when  breaking supersymmetry in an extra dimension. We 
remind that the fermion (gravitino or gaugino)  soft masses can be tuned to be of Dirac-type. Finally,  we briefly connect the 
latter to the study of models with Dirac-type gaugino masses and stress the advantage of having both an F and a D-term sizable contributions 
for the hierarchies of soft-terms as well as for minimizing R-symmetry breaking.
}
%
%\maketitle
\end{titlepage}
\section{Introduction}
\label{intro}

Interest in supersymmetry (SUSY) can be motivated through different arguments among which its role as
 an essential ingredient of the fundamental theory unifying gravity with
all the other interactions in a consistent quantum theory.  At experimentally probed energies,
supersymmetry is not manifest and we would like to think of that as a consequence
of its spontaneous breaking  at a higher energy scale. The spontaneous breaking gives rise in the global limit 
to a \textit{massless} Goldstone fermion, the \textit{goldstino} \cite{Fayet:1974jb,Volkov:1973ix}. Once
gravitational interactions are taken into account, this state is absorbed by 
the gravitino to become the spin-$\frac{1}{2}$ component \cite{Volkov:1973jd,  Fayet:1977vd}  
of the {massive} spin-$\frac{3}{2}$ particle.  The corresponding dynamics is described by the 
Rarita-Schwinger Lagrangian \cite{Rarita:1941mf} which appears supplemented with appropriate constraints. 
We shall review here a few aspects when departing from the minimal set-up.

In the first part, we shall construct a Lagrangian that allows to describe the propagation of a spin-{$\frac{3}{2}$} state in a fluid \cite{Benakli:2013ava}..  
This is obtained as the result of super-Higgs mechanism when supersymmetry is broken by a non-vanishing energy-momentum tensor. The modification 
to the Rarita-Schwinger Lagrangian appears as a deformation of the quadratic mass term. This allows to describe different propagation velocities 
of the different gravitino helicities. The second part reviews the omnipresence of pseudo-goldstinos in models where
supersymmetry is broken in different sectors of models with one extra dimension \cite{Benakli:2007zza}. We restrict to the simplest case of 
at most two branes while the more general
case with an arbitrary number  of supersymmetry breaking sectors and more dimensions can be found in \cite{Benakli:2010nn}. Finally, the last part discusses models 
Dirac gaugino masses (see \cite{Benakli:2011vb} for a review) where the soft terms are induced by gauge mediation. In particular, 
we focus on the advantages of having sizable F and D-term
supersymmetry breaking and point out how this can help to keep R-symmetry unbroken by generating a Dirac gravitino mass.

\section{A Lagrangian for a spin-{$\frac{3}{2}$} propagating in a fluid}
\label{sec-1}

We start by a brief review of the well known non-linear realization of supersymmetry and 
super-Higgs effect.

\subsection{The Rarita-Schwinger Lagrangian from the super-Higgs mechanism}
\label{sec-RS}

\subsubsection{The Volkov-Akulov Lagrangian}
\label{subsubsec-VA}

We consider a global supersymmetric theory in flat space time. Supersymmetry is broken
spontaneously when the vacuum has non-zero energy which, as we will see below, is not the case in
 local supersymmetric theory. If one insists on preserving Lorentz invariance, this is 
accomplished for $N=1$ supersymmetry by giving a vev to an auxiliary field 
in a chiral multiplet (F-term) or in a vector multiplet (D-terms). Without loss of generality,
we shall chose to focus on the F-term case in this section. 
As a consequence of Goldstone theorem,  the low energy spectrum contains a
fermionic massless mode, known as the $\mathit{goldstino}$ for each broken supersymmetry.

The goldstino is a spin $\frac{1}{2}$ field $(G_{\alpha},\bar{G}_{\a})$ in the
$(\frac{1}{2},0)\oplus (0,\frac{1}{2})$ representation of the Lorentz group\footnote{We will
work in 4 dimensions and we use Wess
and Bagger \cite{WessAndBagger} notations. $\eta_{\mu\nu} = diag(-,+,+,+)$,
$\epsilon^{12}=-\epsilon^{21}=1$. $\zeta_{\alpha}$ is a left Weyl
spinor in the $(\frac{1}{2},0)$ representation. $\bar{\zeta}_{\a}$ is a right
Weyl
spinor in the $(0,\frac{1}{2})$ representation. Complex conjugation exchanges
$SU(2)_L$ and $SU(2)_R$. The complex conjugate of a left Weyl spinor is a right
Weyl spinor. We use $\hbar=1$, $c=1$ and denote by $M_P$  the reduced Planck mass:
 $M_P=\sqrt{\frac{\hbar c}{8\pi \textbf{G}}} \simeq 2.435 \times 10^{18}$ GeV with $\textbf{G}$ 
the Newton constant.}.
It has a mass dimension of $\frac{3}{2}$.
Supersymmetry is non-linearly realized on the field $G$ through
\beq
( \epsilon Q + \bar{\epsilon} \bar{Q}) \times G_{\alpha}(x) = \sqrt{2} F \epsilon_{\al}
- i \frac{1}{ \sqrt{2} F}\left[ {G}(x) \si^{\mu} \bar{\epsilon}
- \epsilon \si^{\mu} \bar{{G}}(x)
\right] \partial_{\mu} {G}_{\al}(x).
\eeq
where the F-term $F$ is taken to be real and has a mass dimension 2, 
and plays the role of the order parameter  of supersymmetry breaking.

The invariant (up to a divergence) non-linear Lagrangian for ${G}$ is given 
by the Volkov--Akulov Lagrangian \cite{Volkov:1973ix}
\bea
\scr{L}_{\rm A-V} &=& -F^2 \det \left( \delta_\mu^\nu 
+ i \frac{1}{  F^2} \bar{G}\bar{\sigma}^{\nu}\partial_{\mu}G \right) \\ 
&=& -F^2 - i \bar{G}\bar{\sigma}^{\mu}\partial_{\mu}G 
+ \cdots,
\eea
where the dots refer to higher order terms that we do not discuss here. This canonically 
normalized goldstino field  satisfies the Dirac equation
\bea
\bar{\sigma}^{\mu}\partial_{\mu}G = 0,~~~~ \sigma^{\mu}\partial_{\mu}\bar{G} = 0
\ .
\label{feqgold}
\eea

\subsubsection{The Rarita-Schwinger Lagrangian for a massless gravitino}
\label{masslessRS}

We are interested by  theories with $N=1$ local supersymmetry. The supersymmetric partner of the graviton
is  a gravitino  field
$(\psi_{\mu\alpha},\bar{\psi}_{\mu\a})$ of spin $\frac{3}{2}$ and mass dimension
 $\frac{3}{2}$. Following Fierz and Pauli, the irreducible spin  $\frac{3}{2}$ 
representation is obtained from 
$\psi_{\mu\alpha}$  in the $(\frac{1}{2},\frac{1}{2})\otimes(\frac{1}{2},0) =
(1,\frac{1}{2})\oplus(0,\frac{1}{2})$ representation, and  $\bar{\psi}_{\mu\a}$ 
in the $(\frac{1}{2},\frac{1}{2})\otimes(0,\frac{1}{2}) =
(\frac{1}{2},1)\oplus(\frac{1}{2},0)$ representation by imposing constraints
that project out  the additional spin $\frac{1}{2}$  components. The
$(0,\frac{1}{2})$ and $(\frac{1}{2},0)$ parts in the decomposition of
$(\psi_{\mu\alpha},\bar{\psi}_{\mu\a})$
are removed by imposing
\bea
\bar{\sigma}^{\mu}\psi_{\mu} = 0,~~~~ \sigma^{\mu}\bar{\psi}_{\mu} = 0 \ .
\label{constraint}
\eea
The representations $(1,\frac{1}{2})$ and  $(\frac{1}{2},1)$ have dimension six
each. In order to reduce the number of degrees of freedom
to four we impose
\bea
\partial^{\mu}\psi_{\mu\alpha}=0,~~~~ \partial^{\mu}\bar{\psi}_{\mu\a}=0 \ .
\label{constraintder}
\eea

One can get this structure of equations and constraints from a Lagrangian.
The massless gravitino Rarita-Schwinger Lagrangian is:
\bea
{\cal L}_{\psi} = 
\ep^{\mu\nu\rho\sigma}\bar{\psi}_{\mu}\bar{\sigma}_{\nu}\p_{\rho}\psi_{\sigma} \
.
\eea
The field equations are
\bea
\ep^{\mu\nu\rho\sigma}\bar{\sigma}_{\nu}\p_{\rho}\psi_{\sigma} = 0,~~~~
\ep^{\mu\nu\rho\sigma}\sigma_{\nu}\p_{\rho}\bar{\psi}_{\sigma} = 0 \ .
\eea

By imposing on this equation the condition (\ref{constraint}) we get
\bea
\bar{\sigma}^{\rho}\p_{\rho}\psi_{\sigma} = 0,~~~~
\sigma^{\rho}\p_{\rho}\bar{\psi}_{\sigma} = 0 \ .
\label{gravitinoeq}
\eea
It is easy to see that (\ref{gravitinoeq}) and (\ref{constraint}) imply
(\ref{constraintder}).

\subsubsection{The super-Higgs mechanism and the massive gravitino }
\label{superH-RS}

We would like to promote the non-linear realization of supersymmetry above from a global to a local realization.
The graviton degrees of freedom are described by the vierbein fields $e_\mu^a$ where $a$ is a tangent space index.
We define $e \equiv \det (e_\mu^a)$. The spontaneous F-term supersymmetry breaking is associated with a stress-energy tensor with a vev 
$T_S^{\mu\nu}  = - F^2 \eta^{\mu\nu}$, where $F$ is the vev of the auxiliary field as defined above. 

Promoting $\epsilon$ to a local parameter $\epsilon(x)$, at leading order, the supersymmetry transformation read
\bea
\label{susytran}
\delta e_\mu^a &=& -\frac{1}{M_P}\left(i \bar{\epsilon}\bar{\sigma}^{a}\psi_{\mu} 
- i \epsilon {\sigma}^{a}\bar{\psi}_{\mu}\right) , \nonumber \\
\delta \psi_{\mu} &=& -M_p2\partial_{\mu}\epsilon, \nonumber \\
\delta G &=& \sqrt{2}F \ep \  , \nonumber \\
\delta \bar{\psi}_{\mu} &=& -M_p2\partial_{\mu}\bep , \nonumber \\
\delta \bar{G} &=& \sqrt{2}F\bep \ .
\eea
and, up to a divergence, the supersymmetric Lagrangian is:
\bea
{\cal{L}} = - \frac{1}{2 M_P^2} e R - \ep^{\mu\nu\rho\sigma}\bar{\psi}_{\mu}\bar{\sigma}_{\nu}\partial_{\rho}\psi_{\sigma} 
-F^2 e - i \bar{G}\bar{\sigma}^{\mu}\partial_{\mu}G  -i \frac{ F}{\sqrt{2}M_p}
\left(\psi^{\mu}\sigma_{\mu}\bar{G} +
\bar{\psi}^{\mu}\bar{\sigma}_{\mu}G \right)+ \cdots
\eea
where one sees that the term $F^4$ represents now a cosmological constant. This is problematic as we wish to 
work in a flat background. This issue is solved by adding a combination of a canceling contribution to the 
cosmological constant and a gravitino mass term (where the an anti-symmetric structure  ${\sigma}^{\mu\nu}$ is 
necessary to avoid the appearance of a pathological fermionic term of the form $(\partial G)^2$)  :
\bea
\Delta {\cal{L}}= F^2 e -
m_{\frac{3}{2}}\psi_{\mu}\sigma^{\mu\nu}\psi_{\nu} -
m^*_{\frac{3}{2}}\bar{\psi}_{\mu}{\sigma}^{\mu\nu}\bar{\psi}_{\nu}
-m_{\frac{3}{2}}GG  -  {m^*_{\frac{3}{2}}}\bar{G}\bar{G}\, .
\label{mass}
\eea
and the total Lagrangian is invariant under supersymmetry variations above with the modification:
\bea
%\begin{split}
\label{susyL}
\delta \psi_{\mu} &=&-M_p\left(2\partial_{\mu}\epsilon -i {m^*_{\frac{3}{2}}}
\sigma_{\mu}\bar{\epsilon}\right),~~~~\\ 
\delta \bar{\psi}_{\mu} &=& -M_p\left(2\partial_{\mu}\bep +
 i m_{\frac{3}{2}} \bar{\sigma}_{\mu}\epsilon \right) \ .
%\end{split}
\eea
only if:
\beq
m_{\frac{3}{2}} = \frac{F}{\sqrt{3}M_p} \ .
\label{gravitinomass}
\eeq

Finally, we can go to the unitary gauge by performing the transformation
\bea\label{Unitary}
\psi_{\mu \boldsymbol{\alpha}} \rightarrow \psi_{\mu \boldsymbol{\alpha}}  + 
\frac{\sqrt{2}M_P}{F} \p_\mu G_{\boldsymbol{\alpha}} + i  
\frac{1}{\sqrt{6}} 
\sigma_{\mu \boldsymbol{\alpha \dot\alpha}} \bar G^{\boldsymbol{\dot\alpha}}  \,.
\eea 
and derive the Rarita-Schwinger Lagrangian for a massive gravitino 
\beq
{\cal L}_{g} = 
\ep^{\mu\nu\rho\sigma}\bar{\psi}_{\mu}\bar{\sigma}_{\nu}\p_{\rho}\psi_{\sigma} 
 - m_{\frac{3}{2}}\psi_{\mu}\sigma^{\mu\nu}\psi_{\nu} -
m^*_{\frac{3}{2}}\bar{\psi}_{\mu} {\sigma}^{\mu\nu}\bar{\psi}_{\nu}  \
 .
\eeq

\subsection{Modified Rarita-Schwinger Lagrangian from the super-Higgs mechanism in fluids}

In this section we will derive a Generalized Rarita-Schwinger Lagrangian from
the study the super-Higgs mechanism in fluids. More precisely, we generalize the previous section 
as now supersymmetry is broken spontaneously by the vacuum expectation value of the stress-energy tensor
$T_{\mu\nu}$ which also, in general, breaks spontaneously the Lorentz symmetry.

\subsubsection{The goldstino in supersymmetric fluids: the phonino}\label{fluid}

 Consider for simplicity a supersymmetric field theory in thermal equilibrium described by a background stress-energy tensor
taken to be a perfect fluid:
\begin{equation}
T^{\mu\nu}={\rm diag}\,(\varepsilon,p,p,p) \ .
\label{T}
\end{equation}
where $p$ is the pressure and $\varepsilon$ is the energy density.
This expectation value of the stress-energy tensor (\ref{T}) breaks spontaneously supersymmetry and Lorentz symmetry 
but keeps rotational invariance.

Based on the study of the supersymmetric Ward-Takahashi identity, it was argued
that the associated spontaneous breaking of supersymmetry implies 
 a massless fermionic field in the spectrum, the goldstino  
called here a phonino \cite{Boyanovsky:1983tu}. In fact, the
supersymmetric Ward-Takahashi identity for the supercurrent two-point function:
\begin{equation}\label{WTsusy}
\partial_{\mu}\langle T \{S^{\mu}(x) \bar{S}^{\nu}(y)\}\rangle \sim
\delta^{(4)}(x-y)\langle T^{\nu\rho}\rangle \sigma_{\rho}  \ .
\end{equation}
shows that the
correlator has to have a singularity when $k \to 0$ when going to momentum space 
and assuming a constant energy-momentum tensor the correlator. Note that
without Lorentz invariance it is possible to have a singularity 
without having a massless
particle. While this is known to happen for instance in a free theory, in a generic interacting system 
it is expected that the massless mode is present
(see for example \cite{Kratzert:2002gh} ). Here, we will consider such a situation.

The field equations of the phonino take the form
\beq
T^{\mu\nu}\bar{\sigma}_{\mu}\partial_{\nu}G = 0,~~~~
T^{\mu\nu}\sigma_{\mu}\partial_{\nu}\bar{G} = 0 \ .
\label{Tfeqgold}
\eeq
which can be obtained  from the Lagrangian
\beq
{\cal L}_{G} = -\frac{i}{{\cal T}^{4}}T^{\mu\nu}
\bar{G}\bar{\sigma}_{\mu}\partial_{\nu}G  \ ,
\label{eqGT}
\eeq
where ${\cal T} = | \Tr \, \langle T^{\mu\nu} \rangle |^{\frac{1}{4}}$ has dimension of mass.
Note that for $T^{\mu\nu} = -F^2 \eta^{\mu\nu}$ the Lagrangian (\ref{eqGT}) reduces to
the previous section and the propagator of the phonino becomes that of the usual goldstino.
Note that the gravitino and the goldstino remain massless
in a CFT fluid.

\subsubsection{Modified Rarita-Schwinger Lagrangian }

In the following we will be working at the quadratic order, dropping in particular the four-fermion 
interaction in supergravity, and keep the lowest order of an expansion in powers of 
the dimensionless parameter $\frac {\cal T}{M_p}$. 

The goldstino  variation needs to modified :
\beq
\delta G_{\alpha}(x) = \sqrt{2}\frac{1}{{\cal T}^{2}} g_{\mu\nu}T^{\mu\nu} \epsilon_{\al}
%- i \frac{1}{ \sqrt{2} {\cal T}^{2}} T^{\mu\nu}\left[ {G}(x) \si_{\mu} \bar{\epsilon}
%- \epsilon \si_{\mu} \bar{{G}}(x)
%\right] \partial_{\nu} {G}_{\al}(x) 
+ \cdots,
\eeq
as well as the Lagrangian 
\bea
\scr{L} = -{\cal T}^{4} - i  \frac{T^{\mu\nu}}{  {\cal T}^{4}}\bar{G}\bar{\sigma}_{\mu}\partial_{\nu}G 
+ \cdots,
\eea
As in the usual case, promoting the supersymmerty transformation to local ones requires dealing with 
the contribution of the goldstino energy density to the energy-momentum stress energy tensor. This 
requires adding a canceling cosmological constant term and a gravitino quadratic ``mass'' term. However, as the 
dispersion relation for the phonino is no more Lorentz invariant, we need to allow these
quadratic terms to be non-Lorentz invariant. It is straightforward to see that the the Lagrangian:

\bea\label{finalL}
{\cal {L}} &=& \epsilon^{\mu\nu\rho\sigma} \bar \psi_\mu \bar \sigma_\nu \p_\rho
\psi_\sigma  + \frac{i}{4}  \epsilon^{\mu\nu\rho\sigma}  n_{\sigma \gamma} 
\bar\psi_\mu  \bar \sigma_\rho   \sigma^\gamma
\bar\psi_\nu - \frac{i}{4}  \epsilon^{\mu\nu\rho\sigma}  n_{\sigma \gamma} 
\psi_\mu  \sigma_\rho   \bar \sigma^\gamma
\psi_\nu \nonumber \\
 &-& \frac{i}{\sqrt{2}}  \frac{{\cal T}^2}{M_P }  \frac{T^{\mu\nu}}{ {\cal T}^4}
( \bar
\psi_\mu \bar \sigma_\nu G + \psi_\mu \sigma_\nu \bar G) \nonumber\\
 &+& i \frac{T^{\mu\nu}}{ {\cal T}^4}  \bar G \bar \sigma_\mu \p_\nu G + 
\frac{1}{4} \frac{T^{\mu\nu} n_{\mu\nu}} {{\cal T}^4}G  G +  \frac{1}{4}
\frac{T^{\mu\nu} n_{\mu\nu}} {{\cal T}^4}\bar G  \bar G  \,.\nonumber 
\eea 
is invariant under the modified supersymmetry transformations with Lorentz
violating coefficients:
\bea \label{susytransf}
\delta G^{\boldsymbol{\alpha}} &=& \sqrt{2} {\cal T}^2 \ee^{\boldsymbol{\alpha}} \ ,
\nonumber \\
\delta \psi_{\mu \boldsymbol{\alpha}} &=& - M_P ( 2 \p_\mu
\ee_{\boldsymbol{\alpha}} +i  n_{\mu\nu}
\sigma^\nu_{\boldsymbol{\alpha \dot\alpha}} \bar \ee^{\boldsymbol{\dot\alpha}} ) \ , 
 \\
\delta \bar\psi_{\mu \boldsymbol{\dot\alpha}} &=& - M_P ( 2 \p_\mu
\bar\ee_{\boldsymbol{\dot\alpha}} -i 
n^*_{\mu\nu} \ee^{\boldsymbol{\alpha}} \sigma^\nu_{\boldsymbol{\alpha
\dot\alpha}}  )  \ .\nonumber \
\eea
and leads to the equation of motion for the goldstino. For $n_{\mu\nu}$ real, the requirement
of supersymmetry in flat space implies that it should satisfy
\beq
-\half \epsilon^{\mu\nu\sigma\rho} \epsilon_{\rho}^{~\lambda\gamma\kappa}
n_{\nu\lambda} n_{\sigma\gamma} = 
\frac{T^{\mu\kappa}}{M_P^2} \ .
\label{nT}
\eeq

The unitary gauge is obtained by making a supersymmetry transformation to set $G=0$: 
\bea\label{Unitary2}
\psi_{\mu \boldsymbol{\alpha}} \rightarrow \psi_{\mu \boldsymbol{\alpha}}  + 
\frac{\sqrt{2}M_P}{{\cal T}^2} \p_\mu G_{\boldsymbol{\alpha}} + i  
\frac{M_P}{\sqrt{2}{\cal T}^2}n_{\mu\nu}
\sigma^\nu_{\boldsymbol{\alpha \dot\alpha}} \bar G^{\boldsymbol{\dot\alpha}}  \,.
\eea 

As a result we obtain the Generalized Rarita-Schwinger Lagrangian
\beq
{\cal {L}} = \epsilon^{\mu\nu\rho\sigma} \bar \psi_\mu \bar \sigma_\nu \p_\rho
\psi_\sigma  -\frac{i}{2}  \epsilon^{\mu\nu\rho\sigma}  n_{\sigma}^{~\gamma}
\bar \psi_\mu \bar \sigma_{\rho\gamma}
\bar\psi_\nu +\frac{i}{2}  \epsilon^{\mu\nu\rho\sigma}  n_{\sigma}^{~\gamma} 
\psi_\mu  \sigma_{\rho \gamma} \psi_\nu \,.
\label{gravitinoL}
 \eeq 
and the corresponding equation of motion is
\beq\label{EOM}
\epsilon^{\mu\nu\rho\sigma} \bar \sigma_\nu \p_\rho
\psi_\sigma  -\frac{i}{2}  \epsilon^{\mu\nu\rho\sigma}  n_{\sigma \gamma}  
\bar  \sigma_\rho   \sigma^\gamma
\bar\psi_\nu = 0 \,.
 \eeq

As for the usual Rarita-Schwinger case, two constraints are necessary in order to reduce the number of 
degrees of freedom of $\psi_\mu$ to the four that describe a massive gravitino.
The first is obtained by acting on the equation of motion by $n_{\mu \lambda}  \sigma^\lambda $
which gives
\beq
  -\frac{i}{2}  \epsilon^{\mu\nu\rho\sigma} n_{\mu \lambda}  n_{\sigma \gamma}
\sigma^\lambda    \bar  \sigma_\rho   \sigma^\gamma
\bar\psi_\nu = 0 \,.
 \eeq 
 Using the symmetry of $n_{\mu \lambda} $, this can be put in the form:
\beq\label{cs2}
 T^{\mu\nu}  \sigma_\mu \bar\psi_\nu = 0 \ ,
 \eeq 
This replaces the standard F-term breaking constraint $\bar\sigma^\mu \psi^\mu=0$ of the
gravitino.  A procedure similar to  the case of curved space-time \cite{Corley:1998qg}, allows to get a second
constraint  by taking the component $\mu=0$ of (\ref{EOM}). 

We shall illustrate how to apply these constraint in the next subsection.

\subsubsection{Explicit formulae for a perfect fluid }

Here will show how the gravitino mass can be expressed as a function of 
 the fluid variables. We will consider relativistic ideal fluids with stress-energy tensor 
 \begin{equation}
T^{\mu\nu} = (\epsilon+p)u^{\mu}u^{\nu} + p \eta^{\mu\nu} \ ,
\label{ideal}
\end{equation}
where $u^{\mu}$ is the fluid four-velocity $u^{\mu}u_{\mu} = -1$.
In order to solve (\ref{nT}) we parametrize the solution $n_{\mu\nu}$ as
\beq
n_{\mu\nu} = (n_T- n_L) u_\mu u_\nu + n_T \eta_{\mu\nu} \ .
\eeq 
Plugging $n_{\mu\nu}$ and $T_{\mu\nu}$ and solving for $n_T$ and $n_L$ we get 
\begin{equation}\label{solconstr1}
n_T^2 = \frac{\varepsilon} {3 M_P ^2}  \, , \qquad 
-n_T (n_T + 2 n_L) = \frac{p}{M_P^2}  \,,
\end{equation}
hence
\beq \label{solconstr2}
n_L = - n_T \, \left( \frac{\varepsilon + 3 p}{2 \varepsilon} \right) \ .
\eeq 
Note that for F-term breaking, $\varepsilon = -p, n_L =  n_T$.

For constant energy density and pressure, in the fluid rest frame  $n_{\mu}^{\nu} = diag(n_L,n_T,n_T,n_T)$. We introduce the notation
\begin{equation}
\slash\!\!\!\!D = \sigma^\mu
\del_\mu,  ~~~~~\dslash= \ \sigma^i \del_i \ ,
\end{equation}
and
\begin{equation}
\Psi= \bar \sigma^\mu\psi_\mu,~~~~\bar{\chi} = \bar\sigma^i \psi_i,~~~~ 
{\chi} = \sigma^i \bar\psi_i  \ .
\end{equation}

The constraint  (\ref{cs2}) can be used to solve for one of the components 
\bea\label{cons2}
\psi_0 &=& - v \, \sigma_0  {\chi}  \ ,
\eea 
where $v=\left|\frac{p}{\epsilon}\right|$ is the phonino velocity.
The component $\mu=0$ of equation of motion gives the constraint
\bea\label{cons1}
\dslash \bar{\chi} - i n_T {\chi}+ \del\cdot\psi
&=& 0 \ .
\eea
Putting all the constraints together, and with the correct normalisation of ${\psi}_{\frac{1}{2}}\propto \chi$ \cite{Benakli:2014}, leads to  
\bea\label{eqlong}
(\bar \sigma^0 \del_0 - v \, \bar\dslash ) {\psi}_{\frac{1}{2}}  - i \hat m 
\bar{\psi}_{\frac{1}{2}} &=&0 \,.
\eea 
This is the Dirac equation satisfied by the longitudinal spin-1/2 mode  with
mass 
\beq\label{masslong}
\hat m = \frac{n_L+n_T}{2} = \frac{n_T}{4}\left| (1-3v)\right|= \frac{\sqrt{3}}{4 M_P}
\left|\frac{p-\frac{\varepsilon}{3}}{\sqrt{\varepsilon}}\right| \,. 
\eeq
where as the eqs. (\ref{solconstr1}) determine $n_L,n_T$ only up to a sign, we have used this freedom 
in the last equation to have a positive mass $\hat m$. 

The projector on the transverse part of the spinor is 
\bea
\psi_j &=& \psi_j^T - \left( \frac{1}{2} \sigma_j - \frac{k_j \sla{k}}{2 k^2}
\right) {\psi}_{\frac{1}{2}} + \left( \frac{3 k_j}{2 k^2}+ \half
\frac{\sigma_j \sla{\bar k}}{k^2}  \right) \, k\cdot \psi \ .
\eea 
and the transverse part satisfies then the decoupled equation 
\beq \label{eqtrans}
(\bar\sigma^0 \del_0 + \bar \dslash) \psi_j^T +i \hat m \, \bar \psi_j^T = 0 \ .
\eeq
In the fluid,  the gravitino 
has two distinct propagating modes, the longitudinal and the transverse,
with the same mass but different dispersion relations.

\section{ Pseudo-goldstinos and  Dirac gravitinos from extra dimensions}

In the previous section, we have discussed the presence of a goldstino associated breaking 
supersymmetry in the global limit. We then proceeded to coupling it to gravity in order to obtain
the massive gravitino Lagrangian. We shall now discuss 
a different case where the supersymmetry breaking is intimately related 
with the gravity sector: the goldstino is given by a gravitino component along 
an internal dimension. Only in the limit of comparable sizes of supersymmetry breaking and 
compactification scales are the extra dimensions relevant and we shall therefore focus on 
such case. For simplicity, we will illustrate these in the very simple case of one 
single extra-dimension. We will also mention  the possibility to 
tune the parameters to give Dirac type mass for the gravitino.

\subsection{Minimal supergravity in five dimension}
\label{MinimalSUGRA}

For the sake of establishing our notations and to illustrate the main ideas,  
we shall start the discussion with the simplest case of
 a five-dimensional space parametrized by coordinates $( x^{\mu} , x^{5})$ with 
$\mu=0, \cdots, 3$ and $x^5 \equiv y$ parametrizing the interval 
$S^1/\mathbb{Z}_2$. The latter can be constructed as an orbifold from 
the circle of length $2 \pi R$ ($y \sim y + 2 \pi R$) through the
identification $y \sim - y$. We take the theory in the bulk to be the five-dimensional supergravity 
with the minimal on-shell content: the f\"{u}nfbein $e^{A}_{M}$, the gravitino $\Psi_{M I}$
and the  graviphoton $B_{M}$. We focus on the Lagrangian part involving the gravitino and 
drop the terms involving $B_{M}$ as well as the spin connection as we will consider for simplicity
only the case of a flat extra-dimension.

The on-shell Lagrangian is given 
by\footnote{We recall that we use the approximation of dropping in
the Lagrangian  the four-fermions terms and in the supersymmetry transformations  
the three and four-fermions terms.} \cite{Cremmer:1980gs}:
\begin{eqnarray}
    {\cal L}_{SUGRA} &=& e_{5} \bigg \lbrace -\frac{1}{2}R \left( \omega
\right) 
+ \frac{i}{2} \check{\Psi}_{M}^{I} \Gamma^{MNP} \partial_{N} \Psi_{P I}
+ \cdots 
\bigg \rbrace 
\label{Lsugra}
\end{eqnarray}
and the on-shell supersymmetry transformations are 
:
\begin{eqnarray}
\delta e^{A}_{M}  &=& i \check{\Xi}^{I} \Gamma^{A} \Psi_{M I}
\nonumber   \\
\delta \Psi_{M I}  &=& 2 \partial_{M}\Xi_{I} + \cdots 
\label{SusyTransf0}
\end{eqnarray}
where $\Xi$ is the supersymmetry transformation parameter. 
The five-dimensional spinors $\Psi_{M I} $ and $\Xi_{I}$ are symplectic Majorana
spinors. The five-dimensional gravitino $\Psi_{M I}$
will be written using  the two-component Weyl spinors $\psi_{MI}$ as:
\begin{equation}
\Psi_{M 1} = 
\left(\begin{matrix}
\psi_{M 1}\\
\overline{\psi}_{M 2} 
\end{matrix}\right)  ,\qquad
\Psi_{M 2} = 
\left(\begin{matrix}
-\psi_{M 2}\\
\overline{\psi}_{M 1} 
\end{matrix}\right) 
\label{PsilNotation}
\end{equation}
and the supersymmetry transformation parameter as:
\begin{eqnarray}
\Xi_{1} = - \Xi^{2} =
\left(\begin{matrix}
\epsilon_{1}\\
\overline{\epsilon}_{2}
\end{matrix}\right) 
, & \quad &
\Xi_{2} = \Xi^{1} =
\left(\begin{matrix}
- \epsilon_{2}\\
\overline{\epsilon}_{1}
\end{matrix}\right) 
\nonumber \\
\check{\Xi}_{1} = - \check{\Xi}^{2} =
\left(\begin{matrix}
- \epsilon_{1} ,& \overline{\epsilon}_{2}
\end{matrix}\right) 
, & \quad &
\check{\Xi}_{2} = \check{\Xi}^{1} =
\left(\begin{matrix}
\epsilon_{2} ,& \overline{\epsilon}_{1}
\end{matrix}\right) .
\label{SusyParam}
\end{eqnarray}
The on-shell supersymmetry transformations in
two-component spinor notation are given by:
\begin{eqnarray}
\delta e^{a}_{M}  &=& i \left( \epsilon_{1} \sigma^{a}
\overline{\psi}_{M 1}
+ \epsilon_{2} \sigma^{a} \overline{\psi}_{M 2} \right) + h.c.
\nonumber   \\
\delta e^{\hat{5}}_{M}  &=& \epsilon_{2}\psi_{M 1} - \epsilon_{1} \psi_{M 2} 
+ h.c.
\nonumber   \\
\delta \psi_{1 M}  &=& 2 \partial_{M}\epsilon_{1} + \cdots 
\nonumber   \\
\delta \psi_{2M}  &=& 2 \partial_{M}\epsilon_{2} + \cdots
\label{SusyWeylTransf0}
\end{eqnarray}

The fermionic part of the bulk Lagrangian expressed in two-component spinor
notation reads now:
\begin{eqnarray}
{\cal L}_{Fermi} &=& e_{5} \Bigg \lbrace  \frac{1}{2} \epsilon^{\mu \nu \rho
\lambda} \left( 
\overline{\psi}_{1 \mu}\overline{\sigma}_{\nu}\partial_{\rho}\psi_{1 \lambda}
+ \overline{\psi}_{2 \mu}\overline{\sigma}_{\nu}\partial_{\rho}\psi_{2 \lambda}
\right)
    + e^{5}_{\hat{5}} \left(   \psi_{1 \mu} \sigma^{\mu \nu}\partial_{5}\psi_{2 \nu} 
- \psi_{\mu2} \sigma^{\mu \nu}\partial_{5}\psi_{1 \nu}  \right) 
\nonumber    \\
&&
- e^{5}_{\hat{5}} \left( \psi_{15} \sigma^{\mu \nu}\partial_{\mu}\psi_{\nu2} 
- \psi_{25} \sigma^{\mu \nu}\partial_{\mu}\psi_{1 \nu} 
+ \psi_{1 \mu} \sigma^{\mu \nu}\partial_{\nu}\psi_{25} 
- \psi_{2 \mu} \sigma^{\mu \nu}\partial_{\nu}\psi_{15} \right) 
 + h.c.  + \cdots
\Bigg \rbrace 
\label{Lferm}
\end{eqnarray}
where the five-dimensional covariant derivatives 
expressed have been replaced by partial derivatives as we work in a flat metric.

\subsection{SUSY breaking through twisted boundary conditions}
\label{secSS}

We will perform our study in the simplest case with no branes in the bulk other
than the boundary 
ones at $y = 0$ and $y = \pi R$, as it contains all the qualitative features. 

\subsubsection{The twisted boundary conditions fields basis}
\label{secTwist}

Every generic field $\varphi$ has a well defined $\mathbb{Z}_2$ transformation:
\begin{equation}
\mathbb{Z}_{2} : \quad \varphi(y) \rightarrow {\cal P}_{0} \varphi(- y)
\label{Z2Action}
\end{equation}
that allows us to define the orbifold $S^1/\mathbb{Z}_2$ from the original
five-dimensional compactification on $S^1$. Here ${\cal P}_{0}$ is the parity of
the field $\varphi$ which obeys ${\cal
P}_{0}^{2} = 1$.

The Lagrangian (\ref{Lsugra}) and supersymmetry transformations
(\ref{SusyTransf0}) must be invariant under the action of the mapping (\ref{Z2Action}).
A possible choice of parity assignments is 
\begin{equation}
\psi_{1 \mu}(-y) = + \psi_{1 \mu}(y) .
\label{GravitinoParity}
\end{equation}
At the point $y=0$, the other
fields parity transformations are determined from invariance of supersymmetry
transformations under the mapping (\ref{Z2Action}). We must assign a parity ${\cal P}_{\pi}$ 
for each generic field $\varphi$ at the point
$y=\pi R$ which keeps the Lagrangian and 
the supersymmetry transformations invariant
\begin{equation}
\varphi(\pi R + y) = {\cal P}_{\pi} \varphi(\pi R - y) . 
\label{ParityPi}
\end{equation}

We also need to 
impose periodicity condition, we choose to be:
\begin{equation}
\left(\begin{matrix}
\psi_{M 1} (y + 2 \pi R) \\
\psi_{M 2} (y + 2 \pi R)
\end{matrix}\right) 
=
\left(\begin{matrix}
\cos(2 \pi \omega) & \sin(2 \pi \omega) \\
- \sin(2 \pi \omega) & \cos(2 \pi \omega)
\end{matrix}\right) 
\left(\begin{matrix}
\psi_{M 1} (y) \\
\psi_{M 2} (y)
\end{matrix}\right) 
\label{SSTwist}
\end{equation}
which correspond for $\omega \neq 0$ to implement a Scherk-Schwarz
supersymmetry breaking in the bulk \cite{Scherk:1978ta}.  Then, 
invariance of the supersymmetry transformations  under 
the $\mathbb{Z}_{2}$ mapping (\ref{ParityPi}) determines the parities of all
fields. The result is given in table~\ref{ParitiesPi}, 

\begin{table}[htb]
\centering
\caption{Parity assignments for bulk fields at $y=0$ and $y=\pi R$.}
\label{ParitiesPi}      % Give a unique label
% For LaTeX tables you can use
         \begin{tabular}{|l|c|c|c|c|c|}
         \hline
   ${\cal P}_{0} = +1$ & $e^{a}_{\mu}$ & $e^{\hat{5}}_{5}$ & 
$\psi_{1 \mu}$ & $\psi_{25}$ & $\epsilon_{1}$ \\
         \hline
         ${\cal P}_{0} = -1$ & $e^{a}_{5}$ & $e^{\hat{5}}_{\mu}$ & 
$\psi_{2 \mu}$ & $\psi_{15}$ & $\epsilon_{2}$ \\
         \hline
         ${\cal P}_{\pi} = +1$ & $e^{a}_{\mu}$ & $e^{\hat{5}}_{5}$ & 
$\psi_{\mu +}$ & $\psi_{5 +}$ & $\epsilon_{+}$ \\
         \hline
         ${\cal P}_{\pi} = -1$ & $e^{a}_{5}$ & $e^{\hat{5}}_{\mu}$ & 
$\psi_{\mu -}$ & $\psi_{5 -}$ & $\epsilon_{-}$ \\
          \hline
       \end{tabular}
%\vspace*{5cm}  % with the correct table height
\end{table}
where the following definitions have been introduced:
\begin{eqnarray}
\psi_{\mu +} & = &
\cos( \pi \omega) \psi_{\mu 1} - \sin( \pi \omega) \psi_{\mu 2} 
\nonumber \\
\psi_{\mu -} & = &
\sin( \pi \omega) \psi_{\mu 1} + \cos( \pi \omega) \psi_{\mu 2} 
\nonumber \\
\psi_{5 +} & = &
\sin( \pi \omega) \psi_{5 1} + \cos( \pi \omega) \psi_{5 2} 
\nonumber \\
\psi_{5 -} & = &
\cos( \pi \omega) \psi_{5 1} - \sin( \pi \omega) \psi_{5 2}\nonumber  \\
\epsilon_{+} & = &
\cos( \pi \omega) \epsilon_{1} - \sin( \pi \omega) \epsilon_{2} 
\nonumber \\
\epsilon_{-} & = &
\sin( \pi \omega) \epsilon_{1} + \cos( \pi \omega) \epsilon_{2} 
\label{SusyparametersPi}
\end{eqnarray}

\subsubsection{The Periodic fields basis}
\label{secPeriodicBasis}

It is often useful to work in a basis of periodic fields $\tilde{\psi}_{M I}$ (
ie. $\tilde{\psi}_{M I}(x, y + 2 \pi R) = \tilde{\psi}_{M I}(x ,y)$) in contrast
to the multi-valued 
$\psi_{M I}$ used up to now. These are related by the rotation:
\begin{equation}
\left(\begin{matrix}
\psi_{M 1} \\
\psi_{M 2}
\end{matrix}\right) 
=
\left(\begin{matrix}
\cos[f(y) ] & \sin[ f(y) ] \\
- \sin[ f(y) ] & \cos[ f(y)]
\end{matrix}\right) 
\left(\begin{matrix}
\tilde{\psi}_{M 1} \\
\tilde{\psi}_{M 2}
\end{matrix}\right),
\qquad f(y)= \frac{\omega}{R}y 
\label{PsiRotation}
\end{equation}

The supersymmetry breaking mass terms for the gravitinos is then manifest as we
perform this fields transformation in the kinetic terms of the Lagrangian
 to give the action:
\begin{eqnarray}
    S_{Kinetic} &=&  \int^{2 \pi R}_{0} dy \int d^{4}x \bigg\{ \frac{1}{2}
    e_{5} \bigg[ \frac{1}{2} \epsilon^{\mu \nu \rho
\lambda} \left( 
\tilde{\overline{\psi}}_{\mu
1}\overline{\sigma}_{\nu}\partial_{\rho}\tilde{\psi}_{\lambda 1}
+ \tilde{\overline{\psi}}_{\mu
2}\overline{\sigma}_{\nu}\partial_{\rho}\tilde{\psi}_{\lambda 2} \right) 
\nonumber    \\
&&
    + e^{5}_{\hat{5}} \left(   \tilde{\psi}_{\mu 1} \sigma^{\mu
\nu}\partial_{5}\tilde{\psi}_{\nu 2} 
- \tilde{\psi}_{\mu2} \sigma^{\mu \nu}\partial_{5}\tilde{\psi}_{\nu1}  \right)
- 2 e^{5}_{\hat{5}} \left( \tilde{\psi}_{51} \sigma^{\mu
\nu}\partial_{\mu}\tilde{\psi}_{\nu2} 
- \tilde{\psi}_{52} \sigma^{\mu \nu}\partial_{\mu}\tilde{\psi}_{\nu 1} \right) 
\nonumber    \\
&&
- \left( \frac{\omega}{R} 
\right) e^{5}_{\hat{5}} \left(  \tilde{\psi}_{\mu 1} \sigma^{\mu
\nu}\tilde{\psi}_{\nu 1} +  \tilde{\psi}_{\mu 2} \sigma^{\mu
\nu}\tilde{\psi}_{\nu 2}  \right) \bigg]
+ h.c. \bigg\} .
\label{LkineticEqiv}
\end{eqnarray}
with the fields now being periodic.

Going to the new basis requires then the following redefinition for the
supersymmetry transformation parameters,
\begin{equation}
\left(\begin{matrix}
\epsilon_{1} \\
\epsilon_{2}
\end{matrix}\right) 
=
\left(\begin{matrix}
\cos[ f(y) ] & \sin[ f(y) ] \\
- \sin[ f(y) ] & \cos[ f(y) ]
\end{matrix}\right) 
\left(\begin{matrix}
\tilde{\epsilon}_{1} \\
\tilde{\epsilon}_{2}
\end{matrix}\right), 
\label{SusyParamRedefinition}
\end{equation}
and  the supersymmetry transformations take now the form:
\begin{eqnarray}
\delta \tilde{\psi}_{\mu 1}  &=& 2 \partial_{\mu}\tilde{\epsilon}_{1} +  \cdots
\nonumber    \\
\delta \tilde{\psi}_{\mu 2}  &=& 2 \partial_{\mu}\tilde{\epsilon}_{2} ++ \cdots
\nonumber    \\
\delta \tilde{\psi}_{5 1}  &=& 2 \partial_{5}\tilde{\epsilon}_{1} + 2 \frac{df}{dy}
\tilde{\epsilon}_{2} 
+ \cdots
\nonumber    \\
\delta \tilde{\psi}_{5 2}  &=& 2 \partial_{5}\tilde{\epsilon}_{2} - 2 \frac{df}{dy}
\tilde{\epsilon}_{1} 
+ \cdots
\label{SusyTransfPsiRotated}
\end{eqnarray}
where $\cdots$ stand for higher order terms and terms proportional to $F^{MN}$.

It is important to note that the fields $\tilde{\psi}_{5 1}$ and
$\tilde{\psi}_{5 2}$ transforms non linearly 
under supersymmetry transformations: \textit{they are the  Goldstino fields
associated with the supersymmetry breaking in the bulk} and the supersymmetry
breaking is measured by the $\frac{df}{dy}$ therefore by the transformation between the two basis
or stated differently, by the change of the gravitino component preserved 
at each point of the extra dimension.

\subsubsection{The super-Higgs mechanism}
\label{secSuperH-SS}

From now on we drop the $\tilde{ }$ from the fermion symbols and use 
the periodic basis unless stated differently.

Equations  (\ref{SusyTransfPsiRotated}) show that
four fields  $\psi_{15}$ and $\psi_{25}$ transform non linearly
under supersymmetry transformations. These are the goldstinos associated
with breaking of supersymmetry in the bulk that shall be a``absorbed'' by the 
 two gravitinos for the super-Higgs mechanism. 

In order to study further this effect we will concentrate on the bi-linear terms of 
the fermionic fields: $\psi_{1 \mu}$, $\psi_{2 \mu}$, 
$\psi_{15}$, $\psi_{25}$. Some field redefinition are necessary to obtain standard kinetic terms for the fields
$\psi_{5 I}$:
\begin{eqnarray}
\psi_{1 \mu} & \rightarrow & \psi_{1 \mu} + \frac{i}{\sqrt{6}} \sigma_{\mu}
\overline{\psi}_{25}
\nonumber \\
\psi_{2 \mu} & \rightarrow & \psi_{2 \mu} - \frac{i}{\sqrt{6}} \sigma_{\mu}
\overline{\psi}_{15} 
\nonumber \\
\psi_{15} & \rightarrow & \frac{2}{\sqrt{6}} \psi_{15} 
\nonumber \\
\psi_{25} & \rightarrow & \frac{2}{\sqrt{6}} \psi_{25} .
\label{PsiRed}
\end{eqnarray}
This leads to the Lagrangian density:
\begin{eqnarray}
{\cal L} &=& \frac{1}{2} \bigg \lbrace \frac{1}{2} \epsilon^{\mu \nu \rho
\lambda} \left( 
\overline{\psi}_{1 \mu}\overline{\sigma}_{\nu}\partial_{\rho}\psi_{1 \lambda}
+ \overline{\psi}_{2 \mu}\overline{\sigma}_{\nu}\partial_{\rho}\psi_{2 \lambda}
\right) 
+ \psi_{1 \mu} \sigma^{\mu \nu}\partial_{5} \psi_{2 \nu} 
- \psi_{\mu2} \sigma^{\mu \nu}\partial_{5} \psi_{1 \nu} 
\nonumber    \\
&&
- \frac{i}{2} \left( \overline{\psi}_{15} \overline{\sigma}^{\mu} \partial_{\mu}
\psi_{15} 
+  \overline{\psi}_{25} \overline{\sigma}^{\mu} \partial_{\mu} \psi_{25}
\right) 
+ \psi_{15} \partial_{5} \psi_{25} 
- \psi_{25} \partial_{5} \psi_{15} 
\nonumber    \\
&&
- \frac{\omega}{R} \left( \psi_{1 \mu} \sigma^{\mu \nu} \psi_{1 \nu} 
+  \psi_{2 \mu} \sigma^{\mu \nu} \psi_{2 \nu} 
+  \psi_{15} \psi_{15} + \psi_{25} \psi_{25} \right) 
\nonumber    \\ &&
- i\frac{\sqrt{6}}{2} \left[  \partial_{5} \overline{\psi}_{15}
\overline{\sigma}^{\mu} \psi_{1 \mu} 
+  \partial_{5} \overline{\psi}_{25} \overline{\sigma}^{\mu} \psi_{2 \mu} 
+ \frac{\omega}{R} \left( \overline{\psi}_{25} \overline{\sigma}^{\mu}
\psi_{1 \mu} 
- \overline{\psi}_{15} \overline{\sigma}^{\mu} \psi_{2 \mu}  \right) \right] 
\bigg \rbrace
+ h.c. 
\label{Lkm2}
\end{eqnarray}
The first line represents the five-dimensional kinetic term for the four-dimensional gravitinos, 
the second line corresponds to mass terms coming from the propagation in the fifth dimension, the third line 

\begin{eqnarray}
\label{UnitaryG}
\psi_{1 \mu \boldsymbol{\alpha}} & \rightarrow & \psi_{1 \mu \boldsymbol{\alpha}}  + \sqrt {\frac{2}{3} }
  \frac {R}{\omega}   \p_\mu ( \psi_{25 \boldsymbol{\alpha}} + \frac {R}{\omega}   \p_5  \psi_{15 \boldsymbol{\alpha}} ) + i  
\frac{1}{\sqrt{6}}  
\sigma_{\mu \boldsymbol{\alpha \dot\alpha}} ( \bar \psi_{25}^{ \boldsymbol{ \dot\alpha}} + \frac {R}{\omega} \p_5  \bar  \psi_{15}^{ \boldsymbol{ \dot\alpha}} )  \,.\nonumber  \\
\psi_{2 \mu \boldsymbol{\alpha}} & \rightarrow & \psi_{2 \mu \boldsymbol{\alpha}}  + \sqrt {\frac{2}{3} }
  \frac {R}{\omega}   \p_\mu ( \psi_{15 \boldsymbol{\alpha}} - \frac {R}{\omega}   \p_5  \psi_{25 \boldsymbol{\alpha}} ) + i  
\frac{1}{\sqrt{6}}  
\sigma_{\mu \boldsymbol{\alpha \dot\alpha}} ( \bar \psi_{15}^{ \boldsymbol{ \dot\alpha}} -  \frac {R}{\omega} \p_5  \bar  \psi_{25}^{ \boldsymbol{ \dot\alpha}} )  \,.
\end{eqnarray}

It is straightforward to check that this gauge fixing term provides the
cancellation of mixing terms between gravitino and goldstino fields, which is
the aim of our gauge choice :
\begin{eqnarray}
{\cal L} &=& \frac{1}{2} \bigg \lbrace  
\frac{1}{2} \epsilon^{\mu \nu \rho \lambda} \left( 
\overline{\psi}_{1 \mu}\overline{\sigma}_{\nu}\partial_{\rho}\psi_{1 \lambda}
+ \overline{\psi}_{2 \mu}\overline{\sigma}_{\nu}\partial_{\rho}\psi_{2 \lambda}
\right) 
+ \psi_{1 \mu} \sigma^{\mu \nu}\partial_{5} \psi_{2 \nu} 
- \psi_{\mu2} \sigma^{\mu \nu}\partial_{5} \psi_{1 \nu} 
\nonumber    \\
&&
- \frac{\omega}{R} \left( \psi_{1 \mu} \sigma^{\mu \nu} \psi_{1 \nu} 
+  \psi_{2 \mu} \sigma^{\mu \nu} \psi_{2 \nu} 
\right)  \bigg \rbrace
+ h.c. 
\label{LkmGF}
\end{eqnarray}
 In this gauge, the gravitino  propagators have poles at their physical mass and the degrees of
freedom of would-be goldstinos are eliminated, traded for the longitudinal components for the gravitinos, 
through the super-Higgs  mechanism.

The equations  of motion for the gravitinos $\psi_{\mu I}(y)$ in the unitary gauge 
can be extracted from the Lagrangian (\ref{LkmGF}):
\begin{eqnarray}
- \frac{1}{2} \epsilon^{\mu \nu \rho \lambda}
     \sigma_{\nu}\partial_{\rho} \overline{\psi}_{1 \lambda}
+ \sigma^{\mu \nu}\partial_{5} \psi_{2 \nu} 
- \frac{\omega}{R}\sigma^{\mu \nu} \psi_{1 \nu} 
&=& 0 
\nonumber    \\
- \frac{1}{2} \epsilon^{\mu \nu \rho \lambda}
     \sigma_{\nu}\partial_{\rho} \overline{\psi}_{2 \lambda}
- \sigma^{\mu \nu}\partial_{5} \psi_{1 \nu} 
- \frac{\omega}{R}\sigma^{\mu \nu} \psi_{2 \nu}
    &=& 0
\label{PsiEOM}
\end{eqnarray}
Assuming the gravitinos have a \textit{four-dimensional mass} $m_{3/2}$:
\begin{equation}
\epsilon^{\mu \nu \rho \lambda}
\sigma_{\nu}\partial_{\rho} \overline{\psi}_{\lambda I} = 
- 2 m_{3/2} \sigma^{\mu \nu} \psi_{\nu I} 
\label{PsiMassEq}
\end{equation}
their equations of motion can take the form:
\begin{eqnarray}
\partial_{5} \psi_{2 \mu} 
+ \left( m_{3/2} - \frac{\omega}{R} \right) \psi_{1 \mu}
    &=& 0
\nonumber    \\
\partial_{5} \psi_{1 \mu} 
- \left( m_{3/2} - \frac{\omega}{R} \right) \psi_{2 \mu}
    &=& 0
\label{PsiEOM2}
\end{eqnarray}
A solution for the equations (\ref{PsiEOM2}) in the 
interval $0 < y < \pi R$ satisfying the first condition in (\ref{PsiBC}):
\begin{eqnarray}
\psi_{1 \mu}(y) &=&  \cos \left[ \left( m_{3/2} - \frac{\omega}{R}
\right) y\right] \psi_{1 \mu} (0)
    \nonumber   \\
    \psi_{2 \mu}(y) &=& 
- \sin \left[ \left( m_{3/2} - \frac{\omega}{R} \right) y\right] 
\psi_{1 \mu} (0) .
\label{PsiSolution}
\end{eqnarray}
The second condition in (\ref{PsiBC}) is then used to determine the gravitino
mass:
\begin{equation}
m_{3/2} = \frac{\omega}{R}  + \frac{n}{R} , 
\quad n \in \mathbb{Z}
\label{PsiMass0}
\end{equation}

\subsection{Pseudo-goldstinos and brane localized gravitino mass terms }
\label{secPseudoG}

\subsubsection{Gravitino mass}

Matter fields live on branes  localized for instance at particular points
$y=y_n$. Here, we will consider the simplest case where the branes are
localized on the boundaries $y_b = 0 , \pi R$, as it contains all the qualitative features. 
The generalization can be found in . 
 On each of these branes supersymmetry can be locally broken by the matter scalar potential
and a ``would-be-goldstino`` appears localized. The corresponding action can be written 
as:
\begin{eqnarray}
    S &=&  \int^{2 \pi R}_{0} dy \int d^{4}x \left\{ \frac{1}{2} {\cal L}_{BULK}
+ 
{\cal L}_{0} \delta(y)
    +  {\cal L}_{\pi} \delta(y-\pi R) \right\} .
\label{Action}
\end{eqnarray}

There are four fields 
$\psi_{15}$, $\psi_{25}$, $\chi_{0}$ and $\chi_{\pi}$ transform non linearly
under supersymmetry 
transformations. These are the \textit{``local would be goldstinos''} associated
with breaking of supersymmetry in the bulk and in the two branes respectively.
As we have two gravitinos  then two \textit{local would be goldstinos} will be
absorbed in the super-Higgs effect to give mass to the gravitino fields
$\psi_{1 \mu}$ and $\psi_{2 \mu}$, while  two linear combination of 
the fields $\psi_{15}$, $\psi_{25}$, $\chi_{0}$ and $\chi_{\pi}$ remain as
\textit{pseudo-goldstinos}.

The additional bi-linear terms of the fermionic fields: $\psi_{\mu
1}$, $\psi_{2 \mu}$, 
$\psi_{15}$, $\psi_{25}$, $\chi_{0}$ and $\chi_{\pi}$ take the form:
\begin{eqnarray}
\Delta {\cal L} &=& \delta (y) 
\Bigg\lbrace   - \frac{i}{2} \overline{\chi}_{0} \overline{\sigma}^{\mu}
\partial_{\mu} \chi_{0}
- M_{0} \Bigg[ \psi_{1 \mu} \sigma^{\mu \nu} \psi_{1 \nu} 
+ i \frac{\sqrt{6}}{2} \left(  \overline{\chi}_{0} + \overline{\psi}_{25}
\right) \overline{\sigma}^{\mu} \psi_{1 \mu}
\nonumber \\ &&
+ \left( \chi_{0} + \psi_{25} \right) \left( \chi_{0} + \psi_{25} \right) 
\Bigg] \Bigg\rbrace 
+ \delta (y - \pi R) 
\Bigg\lbrace  - \frac{i}{2} \overline{\chi}_{\pi} \overline{\sigma}^{\mu}
\partial_{\mu} \chi_{\pi}
\nonumber \\ &&
- M_{\pi} \Bigg[ \psi_{1 \mu} \sigma^{\mu \nu} \psi_{1 \nu} 
+ i \frac{\sqrt{6}}{2} \left( \overline{\chi}_{\pi} + \overline{\psi}_{25}
\right) \overline{\sigma}^{\mu} \psi_{1 \mu}
+ \left( \chi_{\pi} + \psi_{25} \right) \left( \chi_{\pi} + \psi_{25} \right)
\Bigg] \Bigg\rbrace 
+ h.c. 
\label{Lkm3}
\end{eqnarray}

The modification necessary to fix the unitary gauge  is straightforward and
two would-be goldstinos are eliminated, absorbed to provide the longitudinal
components for the gravitinos, through the super-Higgs  mechanism while two remain in
the spectrum with masses and fields content explicitly given in . The
 gravitino equations of motion in the bulk-branes system.
The equations 
of motion for the gravitinos $\psi_{\mu I}(y)$ are then given by:
\begin{eqnarray}
- \frac{1}{2} \epsilon^{\mu \nu \rho \lambda}
     \sigma_{\nu}\partial_{\rho} \overline{\psi}_{1 \lambda}
+ \sigma^{\mu \nu}\partial_{5} \psi_{2 \nu} 
- \frac{\omega}{R}\sigma^{\mu \nu} \psi_{1 \nu} 
&=&
    2 M_{0} \sigma^{\mu \nu} \psi_{1 \nu}  \delta(y)
+ 2 M_{\pi} \sigma^{\mu \nu} \psi_{1 \nu} \delta(y - \pi R) 
\nonumber    \\
- \frac{1}{2} \epsilon^{\mu \nu \rho \lambda}
     \sigma_{\nu}\partial_{\rho} \overline{\psi}_{2 \lambda}
- \sigma^{\mu \nu}\partial_{5} \psi_{1 \nu} 
- \frac{\omega}{R}\sigma^{\mu \nu} \psi_{2 \nu}
    &=& 0
\label{PsiEOMX}
\end{eqnarray}

Again, assuming the gravitinos have a \textit{four-dimensional mass} $m_{3/2}$:
\begin{equation}
\epsilon^{\mu \nu \rho \lambda}
\sigma_{\nu}\partial_{\rho} \overline{\psi}_{\lambda I} = 
- 2 m_{3/2} \sigma^{\mu \nu} \psi_{\nu I} 
\label{PsiMassEq2}
\end{equation}
the equations of motion become:
\begin{eqnarray}
\partial_{5} \psi_{2 \mu} 
+ \left( m_{3/2} - \frac{\omega}{R} \right) \psi_{1 \mu}
    &=&
    2 M_{0} \psi_{1 \mu} \delta(y)
+ 2 M_{\pi} \psi_{1 \mu} \delta(y - \pi R) 
\nonumber    \\
\partial_{5} \psi_{1 \mu} 
- \left( m_{3/2} - \frac{\omega}{R} \right) \psi_{2 \mu}
    &=& 0
\label{PsiEOMX2}
\end{eqnarray}
Integration of the equations (\ref{PsiEOMX2}) near the points $y = 0$ and $y =
\pi R$, 
taking into account the parity assumptions, leads to the 
following expressions for the discontinuities of the odd gravitino fields:
\begin{eqnarray}
\psi_{2 \mu}(0^{+}) &=& M_{0} \, \, \psi_{1 \mu} (0) = -\psi_{2 \mu}(0^{-})
\nonumber   \\
\psi_{2 \mu}(\pi R^{-}) &=& - M_{\pi} \, \, \psi_{1 \mu} (\pi R) = - \psi_{\mu
2}(\pi R^{+}).
\label{PsiBC}
\end{eqnarray}

A solution for the equations (\ref{PsiEOMX2})
in the 
interval $0 < y < \pi R$ satisfying the first condition in (\ref{PsiBC}):
\begin{eqnarray}
\psi_{1 \mu}(y) &=& \left\lbrace \cos \left[ \left( m_{3/2} - \frac{\omega}{R}
\right) y\right]
+ M_{0} \sin \left[ \left( m_{3/2} - \frac{\omega}{R} \right) y \right] 
\right\rbrace \psi_{1 \mu} (0)
    \nonumber   \\
    \psi_{2 \mu}(y) &=& \left\{ M_{0} \cos \left[ \left( m_{3/2} -
\frac{\omega}{R}
\right) y\right] 
- \sin \left[ \left( m_{3/2} - \frac{\omega}{R} \right) y\right] \right\}
\psi_{1 \mu} (0) .
\label{PsiSolutionX}
\end{eqnarray}
and the second condition in (\ref{PsiBC}) is then used to determine the gravitino
mass:
\begin{equation}
m_{3/2} = \frac{\omega}{R} + \frac{1}{\pi R} \left[ \arctan \left( M_{0}
\right) 
+ \arctan \left( M_{\pi} \right) \right] + \frac{n}{R} , 
\quad n \in \mathbb{Z}
\label{PsiMass}
\end{equation}

\subsubsection{Pseudo-goldstinos}
We will concentrate now on the would-be goldstino fields $\psi_{15}(y)$, 
$\psi_{25}(y)$, $\chi_{0}$ and $\chi_{\pi}$.  In the unitary gauge, 
a stationary action (in order to derive of the equations of motion) is possible 
if:
\begin{eqnarray}
\partial_{5} \psi_{15} + \frac{\omega}{R} \psi_{25}
&=& - 2 \delta (y) M_{0} \left( \chi_{0} + \psi_{25} \right) 
- 2 \delta (y - \pi R) M_{\pi} \left( \chi_{\pi} + \psi_{25} \right)
\nonumber \\
\partial_{5} \psi_{25} - \frac{\omega}{R} \psi_{15} &=& 0 .
\label{UnitGaugeCond} 
\end{eqnarray}
which imply that the fields $\psi_{5 I}(y)$, in the interval 
$0 < y < \pi R$ can be written as:
\begin{eqnarray}
\psi_{15}(y) &=& \frac{1}{\sqrt{\pi R}} \left[ \cos \left( \frac{\omega}{R} y +
\theta\right)  \chi_{1} +
    \sin \left( \frac{\omega}{R} y + \theta \right) \chi_{2} \right] 
\nonumber \\
\psi_{25}(y) &=& \frac{1}{\sqrt{\pi R}} \left[ \sin \left( \frac{\omega}{R} y +
\theta\right)  \chi_{1} - 
\cos \left( \frac{\omega}{R} y + \theta \right) \chi_{2} \right] 
\label{Psi5UG}
\end{eqnarray}
where $\chi_{1}$ and $\chi_{2}$ are $y$ independent 4d spinors and $\theta$ is a
constant which corresponds to 
a choice of basis for $\chi_{1}$ and $\chi_{2}$.

Integrating the equations (\ref{UnitGaugeCond}) near $y = 0$ and $y = \pi$ we find:
\begin{eqnarray}
\psi_{15}(0^{+}) + M_{0} \left[ \chi_{0} + \psi_{25}(0) \right] &=& 0
\nonumber \\
\psi_{15}(\pi R^{-}) - M_{\pi} \left[ \chi_{\pi} + \psi_{25}(\pi R) \right]
&=& 0
\label{ChiUG}
\end{eqnarray}
which implies (for $M_\pi \neq 0$ and $M_0 \neq 0$):
\begin{eqnarray}
\chi_{\pi} &=& \frac{1}{\sqrt{\pi R}} \left[ - \sin (\omega \pi + \theta) + 
\frac{1}{M_{\pi}} \cos (\omega \pi + \theta) \right] \chi_{1} 
+ \frac{1}{\sqrt{\pi R}} \left[ \cos (\omega \pi + \theta) + 
\frac{1}{M_{\pi}} \sin (\omega \pi + \theta) \right] \chi_{2}
\nonumber \\
\chi_{0} &=& - \frac{1}{\sqrt{\pi R}} \left[ \sin (\theta) + 
\frac{1}{M_{0}} \cos (\theta) \right] \chi_{1} 
+ \frac{1}{\sqrt{\pi R}} \left[ \cos (\theta) - 
\frac{1}{M_{0}} \sin (\theta) \right] \chi_{2} .
\label{ChiUG2}
\end{eqnarray}

Here we see that there are two equations for four fermion fields, and we see how 
the super Higgs mechanism operates: from the original two 5d and
two 4d degrees of 
freedom ($\psi_{15}(y)$, $\psi_{25}(y)$, $\chi_{0}$ and $\chi_{\pi}$), an
infinity of 
Kaluza-Klein modes is absorbed to give mass to the fields $\psi_{1 \mu}(y)$ and
$\psi_{2 \mu}(y)$ and 
only two degrees of freedom remain in the unitary gauge: the pseudo-goldstinos
$\chi_{1}$ and $\chi_{2}$.

\subsection{Dirac gravitino and R-symmetry}
\label{secDirGrav}

We will restore the explicit dependence on the (reduced) five-dimensional
Planck 
mass $ M_{5} = \kappa^{-1} $. It is related to the four-dimensional Planck 
mass $M_{P}$ by
\begin{equation}
\pi R M_{5}^{3} = M_{P}^{2} .
\label{PlanckMasses}
\end{equation}

The  four-dimensional gravitino mass can be read from (\ref{PsiMass}):
\begin{eqnarray}
m_{3/2} &=& \frac{\omega}{R} + \frac{1}{\pi R} \left[ \arctan \left( \kappa
M_{0}
\right) 
+ \arctan \left(\kappa M_{\pi} \right) \right] + \frac{n}{R} ,
\end{eqnarray}

First consider the case with $M_{0}=M_{\pi}=0$. It is well known \cite{Scherk:1978ta, Antoniadis:1990ew, Antoniadis:2004dt} that the Scherk-Schwarz m
mechanism described above, leads for  $\omega =1/2$ to a tower of  Dirac-type Kaluza-Klein excitation of fermions  the bulk fermions. The 
two modes with $n=0$ and $n=-1$ lead to two degenerate Majorana fermions with mass $1/2R$.  The two states correspond to the two orthogonal 
supersymmetry charges of $N=2$ supergravity. One couples to the boundary at $y=0$ and the other one couples to the one at $y=\pi R$.

The original $N=2$ Lagrangian is invariant under an $SU(2)_{\cal R}$ R-symmetry, under
which the gravitinos $\psi_{M1}$ and $\psi_{M2}$ transform in the representation $\textbf{2}$ of $SU(2)_{\cal R}$. For 
$\omega =1/2$ there is a remanent $U(1)_R$ symmetry left with the Dirac gravitino charged under it. This R-symmetry corresponds to the
exchange of  the two boundaries $y=0 \leftrightarrow y=\pi R$.  For generic $\omega \neq 0, 1/2$ the 
R-symmetry is totally broken

Consider the case where there are also contribution the gravitino mass from
potential on the boundary branes. Looking at the mass formula:
\begin{eqnarray}
m_{3/2} &=& \frac{n+\omega}{R} + \frac{1}{\pi R} \left[ \arctan \left( \kappa
M_{0}
\right) 
+ \arctan \left(\kappa M_{\pi} \right) \right], 
\end{eqnarray}

As it stands the localized masses can shift the value of $\frac{\omega}{\pi}$ to any desired value, it could for example
 be   canceled or shifted to $\omega =1/2$ , by the appropriate  choice of values 
of $M_{0}$ and $M_{\pi}$.  However, we are mainly interested in the case when these localized masses  arise
from dynamics on the boundary branes and can have a four-dimensional description. Then, two remarks are in order.

First, the Scherk-Schwarz twist can not compensate the effects of supersymmetry breaking due to F or D-term dynamics
as explicitly shown in \cite{Benakli:2007zza}.

Second, the mass formula can be approximated
by:
\begin{eqnarray}
m_{3/2} &\simeq& \frac{n+\omega}{R} + \frac{\kappa}{\pi R} \left(  M_{0} + 
M_{\pi}
\right). 
\label{PsiMassApprox}
\end{eqnarray}
and as   $\kappa  M_{b}<< 1$ they cannot have a sizable effect on the numerical value of $\omega$ if this is not already
small. For small values of $\omega$, the lightest gravitino lies far below the Kaluza-Klein tower and a four-dimensional
approximation can be used to study the system within four-dimensional supergravity.

\section{Dirac gauginos}
\label{secDiracGaugino}

A few features of the goldstinos have been exhibited in the previous section: i) there might be 
more than one sector breaking supersymmetry (the bulk and all the branes located inside it or at the boundaries).
ii) only a global description of the model allows to identify which of the linear combination
of the  would-be-goldstinos is the (true) goldstino while the rest, the pseudo-goldstinos, remain as 
matter fermions iii) if one starts with an extended R-symmetry due to the presence of of an extended supersymmetry
in the gravitational sector, it is possible to build
models where the breaking parameters can be tuned to keep a part of
this R-symmetry unbroken with the gravitino having a Dirac mass.

While previous sections focused mainly on the gravitational sector, we would like to discuss 
the important role played by similar features in the case of models with Dirac gauginos 
\cite{Fayet:1978qc,Polchinski:1982an}.

Let us first start with the first point: why would we need two or more would-be-goldstinos?

A gaugino $\lambda_a$ acquire Dirac masses $m_D (\lambda_a \psi_a)$ by coupling to other chiral fermions $\psi_a$ in the adjoint representation.
This gaugino mass is soft and can be seen as originating either from  $U(1)_a$ non-vanishing D-terms or F-term $F_b$. In a gauge mediation type scenario, 
new states are introduced at a mass scale $M_m$ (we consider a single scale for simplicity) to serve as mediators of the breaking, and they couple to the visible 
and secluded sector through gauge couplings of strengths $g$, $g_{mD}a$ and $g_{mFb}$, respectively (see for example \cite{Kribs:2010md} for discussion on gravity mediation).
The exchange of loops of these messengers induces soft masses that scale parametrically as:
\begin{eqnarray}
m_{D} &=&  \frac{g}{\sqrt{2}} \left( \sum_{a} c_{Da\frac{1}{2}}\frac{ g_{mDa}}{8\pi^2} \frac{D_a}{M_m} 
+\sum_{b} c_{Fb\frac{1}{2}} \frac{g_{mFb}}{16\pi^2}\frac{F_b^2}{6 M_m^3}  \right) 
%\nonumber \\
%&=& \sum_{a} m_{DDa} + \sum_{b} m_{DFb}
\label{DTermG}
\end{eqnarray}
where the coefficients $c_{Da\frac{1}{2}}$ and $c_{Fb\frac{1}{2}}$ are calculable model dependent coefficients that take into account summation
on other quantum numbers of the messengers.
% in the D-term contributions  $m_{DDa}$ and the F-term contributions $m_{DFb}$, respectively.

The chiral adjoints, pairing up with the gauginos, have scalar superpartners $\Sigma$ in the adjoint representation of the gauge group. The soft induced masses
 are parametrized as $m_\Sigma^2  \tr(\Sigma^\dagger \Sigma) + \frac{1}{2} B_\Sigma  \tr(\Sigma^2 + (\Sigma^\dagger)^2)$ and are given by
\begin{eqnarray}
m_\Sigma^2 &=&  \sum_{a} g_{mDa}^2 \left( c_{Da\Sigma 2} \frac{1}{96\pi^2} \frac{D_a^2}{M_m^2}  +  c_{Da\Sigma 1} \frac{3D_a}{64\pi^2} \right)
+  \sum_{b} g_{mFb}^2 c_{Fb\Sigma 2} \frac{1}{16\pi^2} \frac{F_b^2}{M_m^2} \nonumber \\
B_\Sigma &=& -2   \sum_{a} c_{Da\Sigma 3} \frac{g_{mDa}^2}{96\pi^2} \frac{D_a^2}{M_m^2} 
-2 \sum_{b}  c_{Fb\Sigma 3} \frac{g_{mFb}^2}{16\pi^2} \frac{F_b^2}{M_m^2}
\label{AdjointDtermMasses}
\end{eqnarray}
As the $B_\Sigma$ contribution
tends to make tachyonic the mass of one of the component of the adjoint scalars, the generation of viable soft mass for the adjoint scalars 
turns out to be not totally trivial. The simplest
interactions between the DG-adjoints and the messengers, as the Yukawa couplings descending from $N=2$
lead to tachyonic masses as the $B_\Sigma$ contribution. Historically, this was
the main reason for abandoning the Dirac gaugino scenario in \cite{Fayet:1978qc}. To avoid such a result, the
required forms of the adjoint-messengers interactions have been fully classified in  \cite{Benakli:2008pg, Benakli:2010gi} (see
also \cite{Csaki:2013fla}) .

The scalar partners of the chiral fermions in the visible sector get leading order contribution in $D/ M_m^2$ and $F/ M_m^2$
to their soft masses at three-loop from D-term and two-loop from the F-term:

\begin{eqnarray}
m_{\tilde{f}}^2 = \sum_{i} C_{\tilde{f}}^i \frac{ (m_{iD})^2 \alpha_i}{\pi} \log \left(\frac{m_{\Sigma_P}^{(i)}}{m_{bD}}\right)^2
+ 2 c_{F0 }\sum_{i,b}  C_{\tilde{f}}^i  \left(\frac{\alpha_i}{4\pi} \right)^2 \frac{F_b^2}{M_m^2}
\end{eqnarray}
where $C_{\tilde{f}}^i$ is the quadratic Casimir of the field $f$ under group $i$.

To allow Dirac gaugino masses generated at the leading order in the supersymmetry breaking parameter, and sufficiently heavy selectrons, 
we shall consider a combination of $D$- and F-term breaking, with both $D$- and F-terms comparable. This will generate a spectrum with masses 
of generic order of magnitude
(a)  gaugino masses $\sim \frac{g_{mD} g}{16\pi^2} \frac{D}{M_m}$ (b) sfermion masses $\sim  \frac{g^2}{16\pi^2} \frac{F}{M_m}$
(c) adjoint scalar masses $\sim \frac{g_{mD}}{4\pi} \frac{D}{M_m},\frac{g_{mF}}{4\pi} \frac{F}{M_m} $.
and we expect the adjoint scalars to be the most massive states.  Explicit models can be found, for example, in \cite{Benakli:2010gi}.

The requirement of using both an F-term and a D-term means that the goldstino will be a linear combination of the $U(1)_D$ gaugino
$\lambda_{Da}$ and of the chiral fermion $\chi_b$ associated with the F-term,
%.  The goldstino  $G$  and the orthogonal linear combination $G_{\bot}$ are given by:
\begin{eqnarray}
 G= \frac{\sum_{a} D_a \lambda_{Da} + \sum_{b} F_b \chi_b}{\sqrt{\sum_{a}D_a^2 + \sum_{b}F_b^2}}.
%\qquad G_{\bot}= \frac{F \lambda_D -D \chi}{\sqrt{D^2 + F^2}}.
\end{eqnarray}
The goldstino is "absorbed" through the super-Higgs mechanism as explained in the previous sections. Its coupling to matter is easy to obtain.
 More interesting is the fate of the fermions orthogonal to $G$. The obvious possibility is that all become
pseudo-goldstinos, therefore remains as one of the massive matter fermions. However another
 possibility is one of them will be absorbed as a true goldstino by another gravitino.  
 
 As   $\lambda_D$ and $\chi$ are not visible sector fields, there is no obstruction to make them part of an $N=2$ sector that
 will also include the gravitinos \cite{Antoniadis:2005em}. For instance, in the case of just two would be goldstinos, associated with one
F and one D-term, one can use the orthogonal combination to $G$ to be absorbed by the the second gravitino. This can be tuned as the couplings of the two gravitinos
 to these fields can appear different, even with opposite sign, as seen in the previous section. In this case, the second gravitino will have  the same 
mass   as the visible sector one. The two degenerate Majorana spin-$3/2$ states will combine to give rise to a Dirac gravitino that preserves an R-symmetry.
Of course,  the Higgs sector still breaks R-symmetry as this seems  necessary to obtain the right size of the Higgs mass. But
the induced  Majorana mass for the gauginos can be kept very small. In summary:
\begin{itemize}

\item A generic Dirac gaugino model build for phenomenological purpose would involve at least two different
supersymmetry breaking sources corresponding to non-vanishing D-term and $R$-preserving F-term. 

\item Each sector of the breaking gives rise to a would-be-goldstino and  therefore there are at least two of them. 

\item The gravitino is chosen to to originate in an extended $N=2$ supersymmetric structure. The two gravitinos could
eat two linear combination of the would-be-goldstinos giving rise to degenerate Majorana masses that combine in an
R-preserving Dirac mass.

\end{itemize}

\section*{Acknowledgement}
I wish to thank Luc Darm\'e, Mark Goodsell, Yaron Oz and Giuseppe Policastro for discussions and collaboration
on different parts of the material presented here. This work  was supported in part by French state funds
managed by the ANR Within the Investissements d'Avenir programme under reference
ANR-11-IDEX-0004-02 and in part by the ERC Higgs@LHC.

%\end{acknowledgement}

%
% BibTeX or Biber users please use (the style is already called in the class, ensure that the "woc.bst" style is in your local directory)
% \bibliography{name or your bibliography database}

\end{document}